\documentclass[aps,amsfonts,amsmath,prd,preprint,nofootinbib,tightenlines,showpacs]{revtex4}
\usepackage{epsf}

\def\be{\begin{equation}}
\def\ee{\end{equation}}
\def\bea{\begin{eqnarray}}
\def\eea{\end{eqnarray}}
\def\bml{\begin{subequations}}
\def\blea{\bml\begin{eqnarray}}
\def\elea{\end{eqnarray}\end{subequations}}

\def\mSe{ce}\def\mSo{se}
\def\mJe{Je}\def\mJo{Jo}
\def\mYe{Ye}\def\mYo{Yo}

\def\mHe{He}\def\mHo{Ho}
\def\mIe{Ie}\def\mIo{Io}
\def\mKe{Ke}\def\mKo{Ko}

\begin{document}

\title{Plate with a hole obeys the averaged null energy condition}

\author{Noah Graham}
\email{ngraham@middlebury.edu} \affiliation{Department of Physics,
Middlebury College, Middlebury, VT  05753}
\author{Ken D.\ Olum}
\email{kdo@cosmos.phy.tufts.edu}
\affiliation{Institute of Cosmology, Department of Physics and
Astronomy, Tufts University, Medford, MA  02155}

\begin{abstract}

The negative energy density of Casimir systems appears to violate
general relativity energy conditions.  However, one cannot test the
averaged null energy condition (ANEC) using standard calculations for
perfectly reflecting plates, because the null geodesic would have to
pass through the plates, where the calculation breaks down.  To avoid
this problem, we compute the contribution to ANEC for a geodesic that
passes through a hole in a single plate.  We consider both 
Dirichlet and Neumann boundary conditions in two and three space
dimensions.  We use a Babinet's principle argument to reduce the problem
to a complementary finite disk correction to the perfect mirror result,
which we then compute using scattering theory in
elliptical and spheroidal coordinates.  In the Dirichlet case, we find
that the positive correction due to the hole overwhelms the negative
contribution of the infinite plate.  In the Neumann case, where the
infinite plate gives a positive contribution, the hole contribution is
smaller in magnitude, so again ANEC is obeyed.  These results can be
extended to the case of two plates in the limits of large and small
hole radii.  This system thus provides another example of a situation
where ANEC turns out to be obeyed when one might expect it to be violated.

\end{abstract}


\maketitle

\section{Introduction}

The standard Casimir calculation of the energy density between a pair
of parallel plates (see for example \cite{Mostepanenko97}) yields a
negative energy density between the plates.  While this result poses
no problem in the calculation of the usual Casimir force, it presents a
puzzle for general relativity.  One can construct a spacetime with
an arbitrary geometry $R_{\lambda\nu}$ simply by constructing the
energy-momentum tensor to solve Einstein's equations
\begin{equation}
T_{\lambda\nu} = \frac{1}{8\pi G}\left( R_{\lambda\nu} - \frac{1}{2}
g_{\lambda\nu}R \right) \,.
\end{equation}
The only way to prevent the appearance of exotic phenomena, such as
closed timelike curves \cite{cpc}, traversable wormholes
\cite{Morris88b}, or superluminal travel \cite{Olum:1998mu}, is to place
restrictions on the allowed energy-momentum tensors $T_{\lambda\nu}$.  
While these conditions are all obeyed in classical physics, the
negative energy density of the quantum Casimir system violates most
such conditions, including the weak energy condition (WEC) and the null energy
condition (NEC), which require that $T_{\lambda\nu} V^\lambda V^\nu \geq 0$
for timelike and null vectors respectively.  A still weaker condition,
which is still strong enough to rule out exotic phenomena
(and to prove singularity theorems \cite{Penrose,Galloway,Roman86,Roman88}),
is that the null energy condition hold only when averaged over a
complete geodesic (ANEC).\footnote{For singularity theorems, the
average is only over the future of the trapped surface.}
Geodesics parallel to the plates obey NEC,
so any candidate for ANEC violation would need to pass through the
plates themselves.  Therefore one cannot test ANEC using the
standard Casimir calculation in which one imposes ideal boundary
conditions, since this calculation is not valid within each plate.
One approach to resolve this question is to model the plate using a
domain wall background \cite{Graham:2002yr,Olum:2002ra}; in this case
the effect of the domain wall modifies the calculation significantly,
so that ANEC is obeyed.  In a number of other examples in which one
might expect to find that ANEC is violated, explicit calculation shows
that it is obeyed \cite{Perlov:2003,Graham:2004}. Other calculations
also show that energy condition violation is more difficult to achieve
in realistic situations than idealized models would suggest
\cite{Sopova:2002cs,Sopova:2005}.  ANEC is also known to be obeyed by
free scalar \cite{Klinkhammer} and electromagnetic
\cite{Folacci:1992xg} fields in flat spacetime. Other works have found
restrictions on energy condition violation in flat space
\cite{Borde01,F&Ro95,F&Ro96b}.

In this paper we consider an alternative modification to the Casimir
problem that one might expect would allow the NEC violation between
the plates to extend to ANEC violation:  we imagine a plate with a
small hole, through which the geodesic can pass without encountering
the material of the plate.  Our primary calculation is for the case of 
a single Dirichlet plate, which also leads to a negative energy
density for minimal coupling.  We consider both two and three
spatial dimensions.  We also give extensions of this result to Neumann
boundaries and to the case of two plates in extreme limits.

\section{Null Energy Condition Outside a Boundary}

Since our geodesic never passes through the material that actually
imposes the boundary, we need only  consider the quantum field $\phi$
in empty space.  The effect of the boundary will be to modify the
normal mode expansion for $\phi$.  We will then integrate
$V^\lambda V^\nu T_{\lambda\nu}$, where $T_{\lambda\nu}$ is the
stress-energy tensor, over the null geodesic $V^\lambda$ perpendicular
to the plate, passing through the center of the hole. The
stress-energy tensor for a minimally-coupled scalar field is
\be
\label{eqn:stressenergytensor}
T_{\lambda\nu}=\partial_\lambda\phi\partial_\nu\phi-\frac{1}{2}\eta_{\lambda\nu}
\left[\partial^\lambda\phi\partial_\lambda\phi\right]\,.
\ee
For a null vector,
$\eta_{\lambda\nu}V^\lambda V^\nu= 0$, so we have
\be
T_{\lambda\nu}V^\lambda V^\nu= \left(V^\alpha\partial_\alpha\phi \right)^2\,.
\ee
For a static system, $T_{0i} = 0$ for $i=1,2,3$.  If we further choose
spatial coordinates in which $T_{ij}$ is diagonal and $V = (1,{\bf v})$, then
\be
T_{\lambda\nu}V^\lambda V^\nu = \dot\phi^2+ \sum_i
\left(v_i\partial_i\phi \right)^2\,.
\ee
Let the center of the hole lie at the origin and let the $z$ axis be
the direction perpendicular to the plate, along which the geodesic
lies.  For that path,
\be\label{eqn:perpendicularANEC}
T_{\lambda\nu}V^\lambda V^\nu =  \dot\phi^2+ \left(\partial_z\phi\right)^2\,.
\ee

\section{Babinet's Principle}

It will greatly simplify the scattering theory techniques we would like to
use in our Casimir calculation \cite{Bordag:1996,
Saharian:2000mw,Graham:2002cas} to be able to consider a boundary condition
in a local region.  Therefore we apply a Babinet's principle argument to
reexpress the result for a plate with a hole in terms of the results when
the boundary condition is applied to an entire plate and when a
complementary boundary condition is applied to a disk.  The former is
well-known, while the latter can be computed using scattering theory in
elliptical or spheroidal coordinates.

\begin{figure}
\begin{center}
\leavevmode
\epsfbox{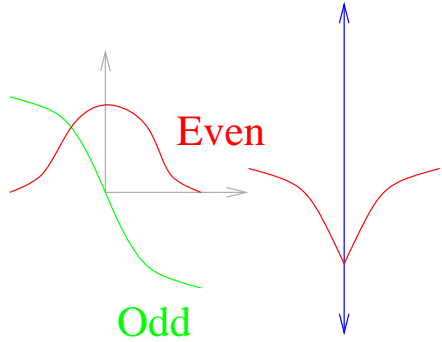}
\end{center}
\caption{\label{fig:Babinet-plate}
In free space (left) we have even
and odd normal mode wave functions.  In the presence of a Dirichlet
plate (right), the even functions are replaced by the odd functions with a
change of sign crossing the plate.}
\end{figure}

We start in empty space, and write the field there in terms of normal
modes that are even or odd in the coordinate across the boundary.
Next we consider a perfectly reflecting Dirichlet boundary with no
holes.  The free-space odd modes obey the boundary conditions, but the
even modes do not.  Instead, we have new even modes, which are just the
odd mode on the right and minus the odd mode on the left, as shown in
Fig.\ \ref{fig:Babinet-plate}.
If we let $E$ denote a sum over the free-space even modes and $O$ the
same sum over the odd modes, then in free space we have $E+O$, whereas
with the barrier we have $O+O$.  Therefore, the renormalized energy,
the difference between the energy with the barrier and the energy in
free space, is $O-E$.  If we have Neumann conditions instead, the
situation is precisely reversed and the energy is $E-O$.

\begin{figure}
\begin{center}
\leavevmode
\epsfbox{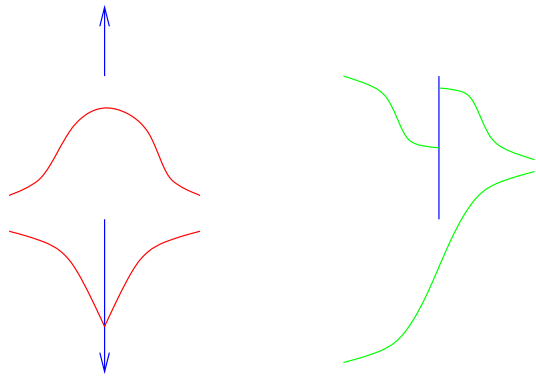}
\end{center}
\caption{\label{fig:Babinet-holes}
If there is a hole in the Dirichlet plate (left), the new even
functions satisfy Neumann conditions in the hole and Dirichlet
conditions elsewhere.  The odd functions for a patch with the same
shape as the hole (right) are the same except for a change of sign
between sides.}
\end{figure}

Now we consider the Dirichlet plate with holes of arbitrary shapes.
Once again the odd modes are unaffected, and
we have new even modes that vanish on the barrier but are
continuous in the holes, as shown in Fig.\ \ref{fig:Babinet-holes}.
Since they are even, they satisfy
Neumann conditions in the hole.  Let us call the contribution of
those modes $A$.  The energy with the perforated barrier is thus
$A+O$, so the renormalized energy is $A-E$.

Finally, suppose that there are Neumann patches where the holes were.  The
even modes are unaffected, but there are new odd modes.  In order to
be odd and continuous they must satisfy Dirichlet conditions on the
plane outside the patches.  Thus, except for a change of sign on one
side, these are the exact same modes of the previous paragraph, as
shown in Fig.\ \ref{fig:Babinet-holes}.  Therefore the total energy is
$A+E$ and the renormalized energy is $A-O$.  Thus we conclude that
\be
\hbox{[Dirichlet plate w/hole]} -
\hbox{[Complementary Neumann disk]} 
= \hbox{[Entire Dirichlet plane]}
\ee
and similarly
\be
\hbox{[Neumann plate w/hole]} - 
\hbox{[Complementary Dirichlet disk]}
= \hbox{[Entire Neumann plane]}.
\ee

\section{Line Segment in Two Spatial dimensions}

In this section we consider a scalar field in 2+1 dimensions with
boundary conditions imposed on a line segment from
$x=-d$ to $x=d$.  In circular coordinates, we can decompose a free, real,
massless scalar field in modes as
\be
\phi(r,\theta)=
\sum_{m=0}^\infty{}'\int_0^\infty dk\,\sqrt{\frac{k}{2\pi\omega}}
J_m(kr)\left( \cos m\theta\, b_{k}^{m}{}^\dagger
+ \sin m\theta\, c_{k}^{m}{}^\dagger\right)
e^{i\omega t}
 + \text{c.c.}
\ee
where the prime on the summation sign indicates that for $m = 0$,
there is no sin mode and instead of $\cos 0 = 1$ we have $1/\sqrt{2}$.

Now we go to elliptical coordinates.  For notational consistency with the
three-dimensional case, we consider the $x$-$z$ plane.  The foci will be
located at $x =\pm d$, and the the geodesic will run along the $z$
axis.  Elliptical coordinates $\mu$, $\theta$ are given by
\bea
x & = & d\cosh\mu\cos\theta\\
z & = & d\sinh\mu\sin\theta
\eea
so
\be
r = \sqrt{x^2+z^2} = 
d\sqrt{\frac{\cosh 2\mu+\cos 2\theta}{2}}\to \frac{d}{2} e^\mu
\ee
as $\mu\to\infty$.

We define our Mathieu functions following the conventions of
Abramowitz and Stegun \cite{Abramowitz} but extending their notation
to be more similar to that of Bessel functions.  The angular functions
are $\mSe_m(\theta, q)$ and $\mSo_m(\theta, q)$.  They satisfy
\be
y'' + (a - 2q\cos 2\theta) y = 0
\ee
where $q=(dk/2)^2$, and are normalized so that
\be
\int_0^{2\pi} d\theta \mSe_m(\theta, q)^2=\int_0^{2\pi} d\theta
\mSo_m(\theta, q)^2 =\pi \,.
\ee
This normalization holds even for $\mSe_0$, but
there is no such function as $\mSo_0$.  Thus the normalization is
precisely the same as the circular functions used above, including the
special case for $m = 0$.

The radial functions of the first kind are $\mJe_m(\mu, q)$ and
$\mJo_m(\mu, q)$ and are precisely the functions called
$Mc_m^{(1)}(\mu, q)$ and $Ms_m^{(1)}(\mu, q)$ respectively in
\cite{Abramowitz}.  They satisfy
\be
y'' - (a - 2q\cosh 2\mu) y = 0
\ee
and go asymptotically to $J_m(\sqrt{q} e^\mu) = J_m (kr)$.   Note
that the functions $Je_m$ and $Jo_m$ defined in
\cite{MorseFeshbach} have an additional factor of $\sqrt{\pi/2}$.
Analogously, we denote the radial functions of the second kind as
$\mYe_m(\mu, q)$ and $\mYo_m(\mu, q)$, which are
$Mc_m^{(2)}(\mu, q)$ and $Ms_m^{(2)}(\mu, q)$ in
\cite{Abramowitz}.

The normalization of radial functions depends only on their
asymptotics, so $\mJe_m$ and $\mJo_m$ have the same normalization as
$J_m$.  Thus the field becomes
\be
\phi(\mu,\theta)=
\sum_{m=0}^\infty{}'\int_0^\infty dk\,\sqrt{\frac{k}{2\pi\omega}}
\left( \mJe_m(\mu, q) \mSe_m (\theta, q) b_{k}^{m}{}^\dagger
+\mJo_m(\mu, q) \mSo_m (\theta, q) c_{k}^{m}{}^\dagger\right)
e^{i\omega t}
 + \text{c.c.}
\ee
where the prime on the summation sign indicates that the second term
is included only for $m > 0$.

Now we consider Neumann conditions along the line segment, which is
$\mu = 0$.  The even wavefunctions obey the conditions already,
because $d\mJe_m/d\mu=0$ at $\mu=0$, but the odd functions need to be
modified.  Instead of $\psi o^m_k= 0$ we need $d\psi o^m_k/d\mu=0$.  For
the free case we had
\be
\mJo_m (\mu, q)=\frac{1}{2}\left[\mHo^{(1)}_m (\mu, q) +
\mHo^{(2)}_m (\mu, q)\right]
\ee
where $\mHo^{(1)} = \mJo_m  + i \mYo_m$ is the function called
$Ms_m^{(3)}$ in \cite{Abramowitz} and 
$\mHo^{(2)}_m= \mJo_m  - i \mYo_m$ is $Ms_m^{(4)}$.  Now we need
\be
\psi o_m(\mu, q)=\frac{1}{2}
\left[e^{2i\delta}\mHo^{(1)}_m (\mu, q)+\mHo^{(2)}_m (\mu, q)\right]
\ee
where
\be
e^{2i\delta}= -\frac{\mHo^{(2)}_m {}' (0, q)}{\mHo^{(1)}_m {}' (0, q)}
\ee
and the derivative is with respect to $\mu$.

We can then compute the renormalized vacuum expectation value of the
time-derivative term in Eq.\ (\ref{eqn:perpendicularANEC}),
\bea
\langle\dot\phi^2\rangle &=&\frac{1}{2\pi}\sum_{m=1}^\infty\int_0^\infty dk\, k
\omega\left(|\psi o_m(\mu, q)|^2-\mJo_m(\mu, q)^2\right)\mSo_m(\mu, q)^2
\eea
and we can write
\be
|\psi o_m(\mu, q)|^2-\mJo_m(\mu, q)^2=
\frac{1}{4}\left[\left(e^{2i\delta}-1\right)\mHo^{(1)}_m
(\mu, q)^2+\left(e^{-2i\delta}-1\right)\mHo^{(2)}_m (\mu, q)^2\right] \,.
\label{eqn:psi}
\ee

We want to extend the range of integration to include negative $k$.
The term $\mSo_m(\mu, q)^2$ is unchanged by going to negative $k$, while $\mJo$
and $\mHo$ behave just like the corresponding Bessel functions.
Thus the situation is exactly as in the circular case
\cite{Saharian:2000mw,Perlov:2003}:  extending the range of
integration exchanges the two terms in Eq.\ (\ref{eqn:psi}), so we can
consider only the first term integrated over the entire real axis.
When we close the contour at infinity, we get the contribution from
the branch cut in $\omega=\sqrt{k}$.  With $k = i\kappa$, the angular
function becomes $\mSo_m (\theta, -q)$, and the radial functions have
the same continuation as Bessel functions,
\bea
\mJo_m(\mu, -q) &=& i^m \mIo_m(\mu, q)\\
\mHo^{(1)}_m(\mu, -q) &=& \frac{2}{\pi} i^{-(m+1)} \mKo_m(\mu, q)
\eea
where $\mIo_m$ and $\mKo_m$ are exactly as in \cite{Abramowitz}.
Thus
\be
e^{2 i \delta_m(i\kappa)}-1
 =(-)^{m+1}i\pi\frac{\mIo_m'(0,\varphi)}{\mKo_m'(0,\varphi)}
\ee
where $\varphi =(d\kappa/2)^2 = -q.$

Putting it all together we have
\bea
\langle\dot\phi^2\rangle&=&
\frac{(-)^m i}{\pi^3}\sum_{m=1}^\infty\int_0^\infty d\kappa
\left(e^{2i\delta}-1\right)\kappa^2 \mKo_m(\mu,\varphi)^2
\mSo_m(\theta,-\varphi)^2\cr
&=&\frac{1}{\pi^2}\sum_{m=1}^\infty\int_0^\infty d\kappa
\frac{\mIo_m'(0,\varphi)}{\mKo_m'(0,\varphi)}
\kappa^2 \mKo_m(\mu,\varphi)^2\mSo_m(\theta,-\varphi)^2 \,.
\eea
On the axis, terms with $m$ even vanish, so we have
\be
\langle\dot\phi^2\rangle =
\frac{1}{\pi^2}\sum_{m=1}^\infty{}' \int_0^\infty d\kappa
\frac{\mIo_m'(0,\varphi)}{\mKo_m'(0,\varphi)}
\kappa^2 \mKo_m(\mu,\varphi)^2\mSo_m(\pi/2,-\varphi)^2
\ee
where the prime on the summation sign indicates that we sum over odd
values of $m$.

The other vacuum expectation value that we need is
$\langle(\partial_z\phi)^2\rangle$.  On the $z$ axis, $\partial_z\phi$
is just the component of the gradient in the $\mu$ direction, which
differs from $\partial_\mu\phi$ by the inverse of the metric
coefficient
\be
h = d\sqrt{\cosh^2\mu-\cos^2\theta} =d\sqrt{\frac{\cosh 2\mu-\cos
2\theta}{2}} \,.
\ee
The calculation is otherwise similar.  Instead of two powers of
$\omega$ from time differentiation we just have the radial function
differentiated with respect to $\mu$,
\be
\langle(\partial_z\phi)^2\rangle=
\frac{1}{h^2}\langle(\nabla_\mu\phi)^2\rangle
=-\frac{1}{\pi^2h^2}\sum_{m=1}^\infty {}' \int_0^\infty d\kappa
\frac{\mIo_m'(0,\varphi)}{\mKo_m'(0,\varphi)}
\mKo_m'(\mu,\varphi)^2\mSo_m(\pi/2,-\varphi)^2\,.
\ee
If instead we have Dirichlet conditions on the line segment, the odd
functions will be unmodified, but for the even functions we need
$\psi e^m_k= 0$ at $\mu = 0$, so we have
\be
\psi e_m(\mu, q)=\frac{1}{2}
\left[e^{2i\delta}\mHe^{(1)}_m (\mu, q)+\mHe^{(2)}_m (\mu, q)\right]
\ee
with
\be
e^{2i\delta}= -\frac{\mHe^{(2)}_m (0, q)}{\mHe^{(1)}_m (0, q)}
\ee
so
\be
e^{2 i \delta_m(i\kappa)}-1
 =(-)^{m+1}i\pi\frac{\mIe_m(0,\varphi)}{\mKe_m(0,\varphi)} \,.
\ee
Thus on the axis we have
\bea
\langle\dot\phi^2\rangle
&=&\frac{1}{\pi^2}\sum_{m=0}^\infty{}^* \int_0^\infty d\kappa
\frac{\mIe_m(0,\varphi)}{\mKe_m(0,\varphi)}
\kappa^2 \mKe_m(\mu,\varphi)^2\mSe_m(\pi/2,-\varphi)^2\\
\langle(\partial_z\phi)^2\rangle
&=&-\frac{1}{\pi^2h^2}\sum_{m=0}^\infty{}^* \int_0^\infty d\kappa
\frac{\mIe_m(0,\varphi)}{\mKe_m(0,\varphi)}
\mKe_m'(\mu,\varphi)^2\mSe_m(\pi/2,-\varphi)^2\,.
\eea
where the star on the summation sign indicates that we sum over even
values of $m$.

\section{Disk in Three Spatial Dimensions}

In this section we consider a scalar field with boundary
conditions imposed on a disk of radius $d$ in the $x$-$y$ plane, centered
at the origin.  In spherical coordinates, we can decompose a free, real,
massless scalar field in modes as
\be
\phi(r,\theta,\phi)=
\sum_{l=0}^\infty\sum_{m=-l}^l
\int_0^\infty dk\,\frac{k}{\sqrt{\pi\omega}} j_l(kr)Y_{lm}(\theta,\phi)
e^{i\omega t}
 + \text{c.c.}
\ee
where $j_l$ is the spherical Bessel function.

Next we go to oblate spheroidal coordinates, given by
\bea
x &=& d\sqrt{(\xi^2+1) (1-\eta^2)}\cos\phi = d \cosh\mu \sin\theta\cos\phi\\
y &=& d\sqrt{(\xi^2+1) (1-\eta^2)}\sin\phi = d \cosh\mu \sin\theta\sin\phi\\
z &=& d\eta\xi = d \sinh\mu \cos \theta
\eea
where $\phi$ is the azimuthal angle, $\eta =\cos\theta$ is the
coordinate akin to polar angle, and $\xi =\sinh\mu$ is the radial
coordinate, with $r = d\xi$ for large $\xi$.

We define the prolate angular spheroidal function $S^m_n(c;\eta) =
S^{m(1)}_n(c;\eta)$ with $c = kd$, using the normalization of Meixner
and Sch\"afke \cite{Meixner}.  Prolate spheroidal functions can be
converted to the oblate ones appropriate to our situation by $k \to
ik$ and $\xi \to -i \xi$.  Thus our angular functions are
$S^m_n(ic;\eta)$, obeying the orthonormality relation
\be
\int_{-1}^1 S^m_n(ic;\eta) S^m_{n'}(ic;\eta)^2d\eta
=\frac{2}{2n+1}\frac{(n+m)!}{(n-m)!}\delta_{nn'}
\ee
Using these functions we can define oblate spheroidal harmonics by analogy
with spherical harmonics,
\def\ynm{{\cal Y}^m_n}
\def\ynz{{\cal Y}^0_n}
\be
\ynm(ic;\eta,\phi) =
\sqrt{\frac{2n+1}{4\pi}\frac{(n-m)!}{(n+m)!}}S^m_n(ic;\eta)e^{im\phi}
\ee
obeying the analogous orthonormality relation
\be
\int_{-1}^1d\eta\int_0^{2\pi} d\phi
\ynm(ic;\eta,\phi)^*{\cal Y}^{m'}_{n'}(ic;\eta,\phi)=\delta_{nn'}\delta_{mm'}
\,.
\ee

We also define the radial spheroidal functions $R^{m(1)}_n(ic;-i\xi)$ and
$R^{m(2)}_n(ic;-i\xi)$, normalized by
\bea
\lim_{\xi\to\infty}R^{m(1)}_n(ic;-i\xi)&=& j_n(c\xi)\\
\lim_{\xi\to\infty}R^{m(2)}_n(ic;-i\xi)&=& y_n(c\xi) \,.
\eea
The radial functions thus have the same normalization as spherical
Bessel functions.  Thus the field becomes
\be
\phi(\xi,\eta,\phi)=
\sum_{n=0}^\infty\sum^n_{m=-n}
\int_0^\infty dk\,\frac{k}{\sqrt{\pi\omega}} \ynm(ic;\eta,\phi)
R^{m(1)}_n(ic;-i\xi)
e^{i\omega t}
 + \text{c.c.}
\ee
If $m+n$ is even, then $S^m_n(ic;\eta)$ is an even function of $\eta$
and $(d/d\xi)R^{m(1)}_n(ic;-i\xi)=0$ at $\xi = 0$.  Thus such wave
functions will be continuous across the disk $\eta = 0$.  Similarly,
if $m+n$ is odd, then $S^m_n(ic;\eta)$ is an odd function of $\eta$
and $R^{m(1)}_n(ic;0)=0$, so the product is once again continuous.
The $R^{(2)}$ functions do not have these boundary conditions, so
they cannot be used in the vacuum wave functions.

Now we consider Neumann conditions on the disk
$\xi = 0$.  If $m+n$ is even, the functions obey the conditions
already.  Otherwise we need to combine $R^{m(1)}_n$ and $R^{m(2)}_n$ to
give the desired condition.  With $R^{m(3)}_n=R^{m(1)}_n+iR^{m(2)}_n$ and 
$R^{m(4)}_n=R^{m(1)}_n-iR^{m(4)}_n$ we can write the desired radial function
\be
\psi^m_n(ic;-i\xi)=\frac{1}{2}
\left[e^{2i\delta(ic)}R^{m(3)}_n(ic;-i\xi) + R^{m(4)}_n(ic;-i\xi)\right]
\ee
with the condition
\be
e^{2i\delta(ic)}= -\frac{R^{m(4)}_n{}'(ic;0)}{R^{m(3)}_n{}'(ic;0)}
\ee
where the derivative is with respect to the second argument.

The vacuum expectation value of the time derivative term, subtracting
the free vacuum, then becomes
\bea
\langle\dot\phi^2\rangle &=&\frac{1}{\pi}
\sum_{n=0}^\infty\sum^n_{m=-n}{}' \int_0^\infty dk\, 
k^2 \omega
|\ynm(ic;\eta,\phi)|^2
\left[|\psi_m(ic;-i\xi)|^2-R^{m(1)}_n(ic;-i\xi)^2\right]\cr
& = &\frac{1}{4\pi}
\sum_{n=0}^\infty\sum^n_{m=-n}{}' \int_0^\infty dk\,
k^2 \omega
|\ynm(ic;\eta,\phi)|^2
\big[\left(e^{2i\delta(ic)}-1\right)R^{m(3)}_n(ic;-i\xi)^2\cr
&&\qquad\qquad\qquad\qquad
+\left(e^{-2i\delta(ic)}-1\right)R^{m(4)}_n(ic;-i\xi)^2\big]
\eea
where the prime on the summation sign means that only odd values of
$m+n$ are included.

We want to extend the range of integration to include negative $k$,
which changes the sign of $c$.  We can implement this change by
changing the sign of $\xi$ in $R^{m(1)}_n$ and $R^{m(2)}_n$, since these
real functions are not affected by complex conjugation.  The functions
$R^{m(1)}_n$ and $R^{m(2)}_n$ have opposite parity under this
transformation, so $R^{m(3)}_n$ and $R^{m(4)}_n$ change places.  Thus
including negative $k$ in the first term gives the second term, and we have
\be
\langle\dot\phi^2\rangle=\frac{1}{4\pi}
\sum_{n=0}^\infty\sum^n_{m=-n}{}' \int_{-\infty}^\infty dk\,
k^2 \omega
|\ynm(ic;\eta,\phi)|^2
\left(e^{2i\delta(ic)}-1\right)R^{m(3)}_n(ic;-i\xi)^2 \,.
\ee

If we take $k$ and thus $c$ in the upper half plane we will get
spheroidal functions whose parameter goes to negative real infinity.
Therefore we can close the contour at infinity and obtain
an integral along the branch cut on the imaginary axis associated with
the square root in  $\omega$,
\bea
\langle\dot\phi^2\rangle &=& \frac{1}{2\pi}
\sum_{n=0}^\infty\sum^n_{m=-n}{}' \int_{0}^\infty d\kappa\,
\kappa^3 |\ynm(\gamma;\eta,\phi)|^2
\left(e^{2i\delta(\gamma)}-1\right)R^{m(3)}_n(\gamma;-i\xi)^2 \cr
&=& -\frac{1}{\pi}
\sum_{n=0}^\infty\sum^n_{m=-n}{}' \int_{0}^\infty d\kappa\,
\kappa^3\frac{R^{m(1)}_n {}'(\gamma;0)}{R^{m(3)}_n {}'(\gamma;0)}
|\ynm(\gamma;\eta,\phi)|^2 R^{m(3)}_n(\gamma;-i\xi)^2
\eea
where $\gamma = ic = ikd = -\kappa d$.  On the axis, we have
\be
\langle\dot\phi^2\rangle =
-\frac{1}{\pi} \sum_{n=1}^\infty{}' \int_{0}^\infty d\kappa\,
\kappa^3\frac{R^{0(1)}_n {}'(\gamma;0)}{R^{0(3)}_n {}'(\gamma;0)}
|\ynz(\gamma;1,\phi)|^2 R^{0(3)}_n(\gamma;-i\xi)^2
\ee
where we have specialized to $m = 0$ because the contributions from
nonzero $m$ vanish on the axis, leaving only a sum over odd values
of $n$.  Similarly, on the axis we have
\be
\langle (\partial_z \phi)^2\rangle =
-\frac{1}{\pi d^2}
\sum_{n=1}^\infty{}' \int_{0}^\infty  d\kappa \,
\kappa \frac{R^{0(1)}_n {}'(\gamma;0)}{R^{0(3)}_n {}'(\gamma;0)}
|\ynz(\gamma;1,\phi)|^2 R^{0(3)}_n{}'(\gamma;-i\xi)^2
\ee
where the primes on the radial functions indicate derivatives with
respect to the second argument, and the metric coefficient 
\be
h_\xi = \left|\frac{\partial\mathbf{x}}{\partial\xi}\right|
=d\sqrt{\frac{\xi^2 + \eta^2}{\xi^2 + 1}} \,.
\ee
becomes equal to $d$ on the axis.

For Dirichlet conditions on the disk, we have the analogous results
\bea
\langle\dot\phi^2\rangle &=& -\frac{1}{\pi}
\sum_{n=0}^\infty{}^* \int_{0}^\infty d\kappa\,
\kappa^3\frac{R^{0(1)}_n(\gamma;0)}{R^{0(3)}_n(\gamma;0)}
|\ynz(\gamma;\eta,\phi)|^2
R^{0(3)}_n(\gamma;-i\xi)^2 \\
\label{eqn:D3axis}
\langle (\partial_z \phi)^2\rangle &=&
- \frac{1}{\pi d^2}
\sum_{n=0}^\infty{}^* \int_{0}^\infty d\kappa \,
\kappa \frac{R^{0(1)}_n(\gamma;0)}{R^{0(3)}_n(\gamma;0)}
|\ynz(\gamma;1,\phi)|^2 R^{0(3)}_n{}'(\gamma;-i\xi)^2
\eea
where we have again specialized to the axis so that only $m = 0$
contributes, the derivatives of the radial functions are again with
respect to the second argument, and the star on the summation sign
indicates that now we sum over even values of $n$.

\section{Numerical Calculation}

For a null geodesic $V^\lambda$ perpendicular to the plate and passing
through the center of the hole, the contribution to ANEC is given by Eq.\
(\ref{eqn:perpendicularANEC}).  We can compute the
results for the complementary disk using the formulae derived in the
previous sections.  We then add the complete plate results,
\be
\langle\dot\phi^2\rangle + \langle(\partial_z \phi)^2\rangle =
\left\{ \begin{array}{l@{\quad\quad}l}
\displaystyle
\mp \frac{1}{32 \pi z^3} & \hbox{in two dimensions, and} \cr
\displaystyle
\mp \frac{1}{16 \pi^2 z^4} & \hbox{in three dimensions}
\end{array}\right.
\ee
with Dirichlet conditions giving the upper sign and Neumann the lower.

In two dimensions, we compute the Mathieu functions using the
package of Alhargan \cite{Alhargan:2000,Alhargan:2000a}, with some
minor modifications:  we use 80-bit double precision throughout
the calculation to accommodate the extreme dynamical range needed for
the wide range of Mathieu function parameters we use, and we have
adapted the code to use our set of normalization conventions.  These C++
routines are then imported into Mathematica, where we can use
efficient routines for numerical sums and integrals.

In three dimensions, we use the Mathematica spheroidal harmonic
package of Falloon \cite{Falloon:2002}.  We have updated it to fix
incompatibilities with the latest version of Mathematica and to avoid
memory leaks.  We have also made a number of efficiency optimizations
appropriate to the unusual demands we make on the code (for example,
we changed the caching structure so that it is appropriate to the way we
call the functions, with the same arguments but different
parameters rather than the other way around; we also wrote specific
code for the modified radial function of the third kind to avoid
cancellations of exponentially growing quantities).

\begin{figure}
\begin{center}
\leavevmode
\epsfbox{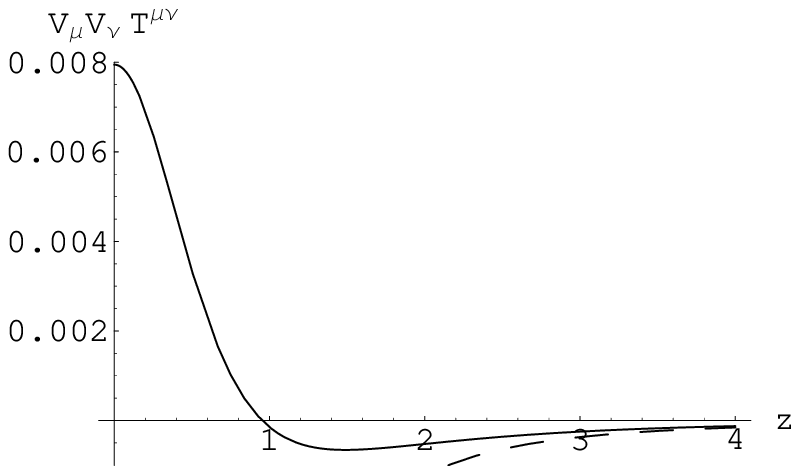}
\epsfbox{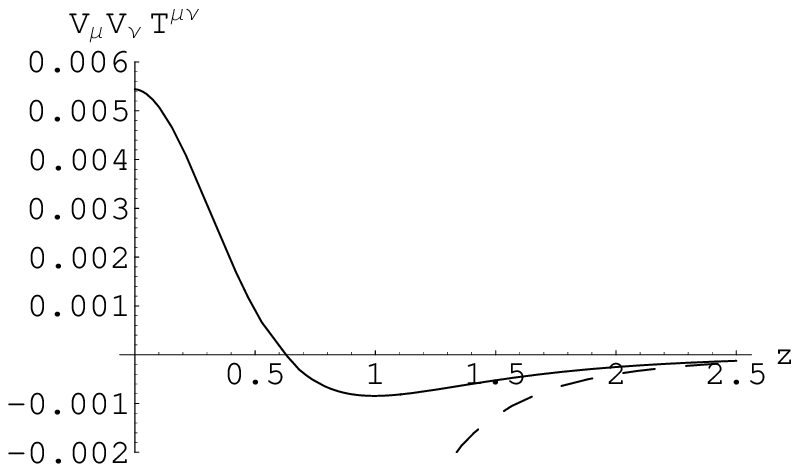}
\end{center}
\caption{
\label{fig:graphs}
Contributions to NEC in two dimensions (left) and three dimensions
(right) for a Dirichlet plate with a hole of unit radius, as functions
of distance along the axis passing through the center of the hole.
Extrapolation is used for points at a distance less than $0.15$ 
in the left panel and $0.25$ in the right panel.  The dotted lines
show the perfect mirror result.
}
\end{figure}

Figure \ref{fig:graphs} shows the contributions to NEC for Dirichlet
plates with holes of unit radius in two and three spatial dimensions,
as functions of distance along the axis.  Using Babinet's principle,
we have computed the sum of contributions from a Neumann disk and an
infinite Dirichlet mirror.  At small distances, the contributions from
both the finite disk and the infinite mirror diverge like $1/z^{n+1}$,
where $z$ is the distance from the origin and $n$ is the spatial dimension.  
The true result, however, does not diverge (the origin is just a point
in empty space) and by symmetry must have zero slope at the origin.
This cancellation provides a highly nontrivial check on our
calculation. Going all the way to the origin would require infinite
precision; in two dimensions we stop at a distance $0.15 d$, while in
three dimensions we stop at distance $0.25 d$.  At these values, our curves
already show this cancellation clearly.  We also extrapolate our
result (without putting in any restrictions on the extrapolation at
the origin) and find that it goes smoothly to a finite value with zero
slope at $z=0$.  A less stringent check is that our calculation
approaches the perfect mirror at large distances; in our approach this
result simply tells us that the finite disk contribution is going to
zero fast enough.

\section{Discussion}
The results in Fig.\ \ref{fig:graphs} are quite striking.  In both
cases, far from the origin we see the negative contribution to ANEC
that we would expect from the standard calculation.  Near the plate,
however, the hole leads to a large positive contribution.  Integrating
the results shown in Fig.\ \ref{fig:graphs} gives a total contribution
(including both sides of the plate) of $1.63 \times 10^{-3}/d^{3}$ in
three dimensions and $4.53 \times 10^{-3}/d^{2}$ in two dimensions.
The positive contribution overwhelms the negative contribution, so
that ANEC is obeyed.

\begin{figure}
\begin{center}
\leavevmode
\epsfbox{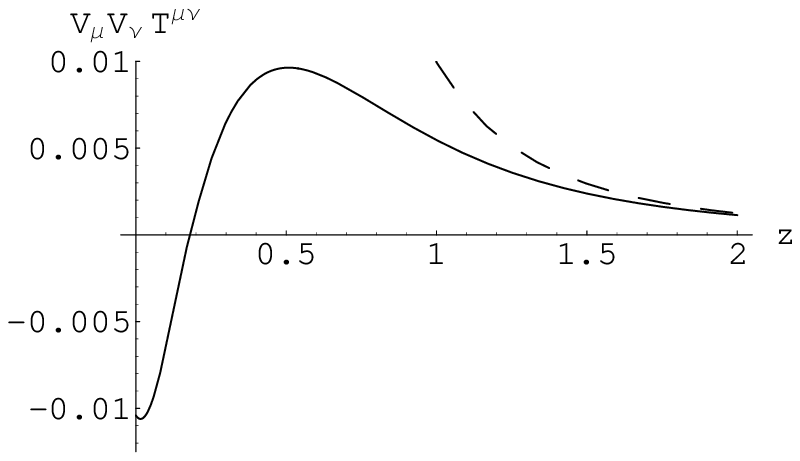}
\epsfbox{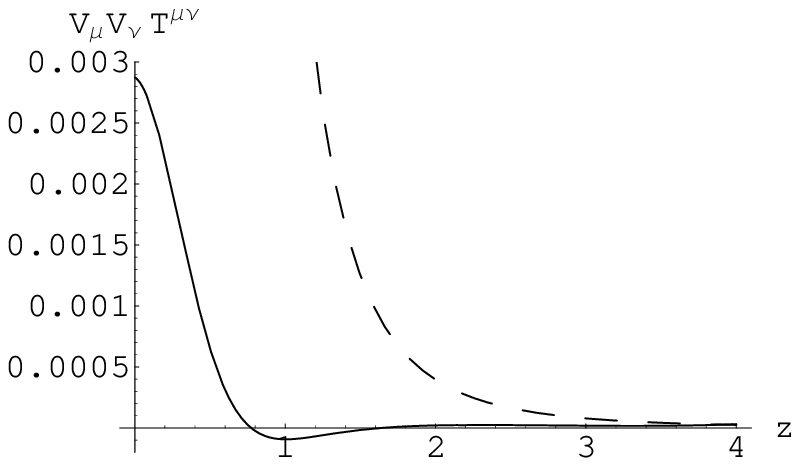}
\end{center}
\caption{
\label{fig:graphsn}
Contributions to NEC in two dimensions (left) and three dimensions
(right) for a Neumann plate with a hole of unit radius, as functions
of distance along the axis passing through the center of the hole.
Extrapolation is used for points at a distance less than $0.11$ 
in the left panel and $0.25$ in the right panel.  The dotted lines
show the perfect mirror results.
}
\end{figure}

We can look at these results from a different point of view by considering
conformal coupling.  Since the NEC contribution in this case differs from
that in minimal coupling by a total derivative, it leads to the same
results for ANEC.  From this point of view, one might not expect any ANEC
violation in the case at hand, because the quantum contributions to the
perfect mirror vanish.  (In the perfect mirror case, changing from minimal
to conformal coupling effectively moves the negative contributions from the
region outside the boundary onto the boundary itself \cite{Ford:1998he}.) 
However, from this point of view one might just as well expect Neumann
conditions to violate ANEC; while the perfect Neumann mirror result is
positive in minimal coupling, it also vanishes for conformal coupling.  The
results for Neumann conditions are shown in Fig.\ \ref{fig:graphsn}. Once
again integration gives positive results, $2.10 \times 10^{-3} / d^{3}$ in
three dimensions and $1.77 \times 10^{-2} / d^{2}$ in two
dimensions, so ANEC is obeyed.  Thus again from this point of view,
one finds ANEC obeyed more often than would be na\"ively expected.

A conformal field between two plates would have a constant negative
energy density.  We can estimate how our results would extend to the
case of parallel plates with holes, in two limits.  If the separation
between the plates $\ell$ is much smaller than the radius of the hole
$d$, the two plates are equivalent to a single plate and ANEC
continues to hold.  In the other extreme, if the separation between
the plates is large compared to the radius of the hole, then we can
assume that the change $\Delta$ in ANEC induced by adding a hole in
one plate is unaffected by the other plate.  In three dimensions, we
obtain a contribution to ANEC of
\be\label{eqn:2plates}
\int dx \, V^\lambda V^\nu T_{\lambda \nu} \approx 2 \Delta -
\frac{\pi^2}{720 \ell^3}
\ee
where the first term is the effect of the each of the two holes
individually and the second term is the standard contribution from the
two plates.  For Dirichlet conditions, $\Delta = 1.63 \times
10^{-3}/d^3$.  Thus Eq.\ (\ref{eqn:2plates}) gives a positive result
as long as $\ell > 1.6d$, which surely includes its entire range of
applicability.  Similar results hold for the other cases we have considered.

\section{Acknowledgments}
K.\ D.\ O. thanks Xavier Siemens for helpful conversations.
N.\ G. thanks the Kavli Institute for Theoretical Physics for
hospitality while part of this work was completed.  K.\ D.\ O. was
supported in part by the National Science Foundation (NSF).  N.\ G.\
was supported in part by the NSF through the Vermont Experimental
Program to Stimulate Competitive Research (VT-EPSCoR).

\bibliographystyle{apsrev}
\bibliography{gr}

\begin{thebibliography}{29}
\expandafter\ifx\csname natexlab\endcsname\relax\def\natexlab#1{#1}\fi
\expandafter\ifx\csname bibnamefont\endcsname\relax
  \def\bibnamefont#1{#1}\fi
\expandafter\ifx\csname bibfnamefont\endcsname\relax
  \def\bibfnamefont#1{#1}\fi
\expandafter\ifx\csname citenamefont\endcsname\relax
  \def\citenamefont#1{#1}\fi
\expandafter\ifx\csname url\endcsname\relax
  \def\url#1{\texttt{#1}}\fi
\expandafter\ifx\csname urlprefix\endcsname\relax\def\urlprefix{URL }\fi
\providecommand{\bibinfo}[2]{#2}
\providecommand{\eprint}[2][]{\url{#2}}

\bibitem[{\citenamefont{Mostepanenko and Trunov}(1997)}]{Mostepanenko97}
\bibinfo{author}{\bibfnamefont{V.~M.} \bibnamefont{Mostepanenko}}
  \bibnamefont{and} \bibinfo{author}{\bibfnamefont{N.~N.}
  \bibnamefont{Trunov}}, \emph{\bibinfo{title}{The Casimir Effect and its
  Applications}} (\bibinfo{publisher}{Clarendon Press},
  \bibinfo{address}{Oxford}, \bibinfo{year}{1997}).

\bibitem[{\citenamefont{Hawking}(1992)}]{cpc}
\bibinfo{author}{\bibfnamefont{S.~W.} \bibnamefont{Hawking}},
  \bibinfo{journal}{Phys. Rev. D} \textbf{\bibinfo{volume}{46}},
  \bibinfo{pages}{603} (\bibinfo{year}{1992}).

\bibitem[{\citenamefont{Morris et~al.}(1988)\citenamefont{Morris, Thorne, and
  Yurtsever}}]{Morris88b}
\bibinfo{author}{\bibfnamefont{M.~S.} \bibnamefont{Morris}},
  \bibinfo{author}{\bibfnamefont{K.~S.} \bibnamefont{Thorne}},
  \bibnamefont{and}
  \bibinfo{author}{\bibfnamefont{U.}~\bibnamefont{Yurtsever}},
  \bibinfo{journal}{Phys. Rev. Lett.} \textbf{\bibinfo{volume}{61}},
  \bibinfo{pages}{1446} (\bibinfo{year}{1988}).

\bibitem[{\citenamefont{Olum}(1998)}]{Olum:1998mu}
\bibinfo{author}{\bibfnamefont{K.~D.} \bibnamefont{Olum}},
  \bibinfo{journal}{Phys. Rev. Lett.} \textbf{\bibinfo{volume}{81}},
  \bibinfo{pages}{3567} (\bibinfo{year}{1998}),
  \eprint[http://arXiv.org/abs]{gr-qc/9805003}.

\bibitem[{\citenamefont{Roman}(1986)}]{Roman86}
\bibinfo{author}{\bibfnamefont{T.~A.} \bibnamefont{Roman}},
  \bibinfo{journal}{Phys. Rev. D} \textbf{\bibinfo{volume}{33}},
  \bibinfo{pages}{3526} (\bibinfo{year}{1986}).

\bibitem[{\citenamefont{Penrose}(1965)}]{Penrose}
\bibinfo{author}{\bibfnamefont{R.}~\bibnamefont{Penrose}},
  \bibinfo{journal}{Phys. Rev. Lett.} \textbf{\bibinfo{volume}{14}},
  \bibinfo{pages}{57} (\bibinfo{year}{1965}).

\bibitem[{\citenamefont{Galloway}(1981)}]{Galloway}
\bibinfo{author}{\bibfnamefont{G.~J.} \bibnamefont{Galloway}},
  \bibinfo{journal}{Manuscr. Math.} \textbf{\bibinfo{volume}{35}},
  \bibinfo{pages}{209} (\bibinfo{year}{1981}).

\bibitem[{\citenamefont{Roman}(1988)}]{Roman88}
\bibinfo{author}{\bibfnamefont{T.~A.} \bibnamefont{Roman}},
  \bibinfo{journal}{Phys. Rev. D} \textbf{\bibinfo{volume}{37}},
  \bibinfo{pages}{546} (\bibinfo{year}{1988}).

\bibitem[{\citenamefont{Graham and Olum}(2003)}]{Graham:2002yr}
\bibinfo{author}{\bibfnamefont{N.}~\bibnamefont{Graham}} \bibnamefont{and}
  \bibinfo{author}{\bibfnamefont{K.~D.} \bibnamefont{Olum}},
  \bibinfo{journal}{Phys. Rev.} \textbf{\bibinfo{volume}{D67}},
  \bibinfo{pages}{085014} (\bibinfo{year}{2003}),
  \eprint[http://arXiv.org/abs]{hep-th/0211244}.

\bibitem[{\citenamefont{Olum and Graham}(2003)}]{Olum:2002ra}
\bibinfo{author}{\bibfnamefont{K.~D.} \bibnamefont{Olum}} \bibnamefont{and}
  \bibinfo{author}{\bibfnamefont{N.}~\bibnamefont{Graham}},
  \bibinfo{journal}{Phys. Lett.} \textbf{\bibinfo{volume}{B554}},
  \bibinfo{pages}{175} (\bibinfo{year}{2003}),
  \eprint[http://arXiv.org/abs]{gr-qc/0205134}.

\bibitem[{\citenamefont{Schwartz-Perlov and Olum}(2003)}]{Perlov:2003}
\bibinfo{author}{\bibfnamefont{D.}~\bibnamefont{Schwartz-Perlov}}
  \bibnamefont{and} \bibinfo{author}{\bibfnamefont{K.~D.} \bibnamefont{Olum}},
  \bibinfo{journal}{Phys. Rev.} \textbf{\bibinfo{volume}{D68}},
  \bibinfo{pages}{065016} (\bibinfo{year}{2003}),
  \eprint[http://arXiv.org/abs]{hep-th/0307067}.

\bibitem[{\citenamefont{Graham et~al.}(2004)\citenamefont{Graham, Olum, and
  Schwartz-Perlov}}]{Graham:2004}
\bibinfo{author}{\bibfnamefont{N.}~\bibnamefont{Graham}},
  \bibinfo{author}{\bibfnamefont{K.~D.} \bibnamefont{Olum}}, \bibnamefont{and}
  \bibinfo{author}{\bibfnamefont{D.}~\bibnamefont{Schwartz-Perlov}},
  \bibinfo{journal}{Phys. Rev.} \textbf{\bibinfo{volume}{D70}},
  \bibinfo{pages}{105019} (\bibinfo{year}{2004}),
  \eprint[http://arXiv.org/abs]{gr-qc/0407006}.

\bibitem[{\citenamefont{Sopova and Ford}(2005)}]{Sopova:2005}
\bibinfo{author}{\bibfnamefont{V.}~\bibnamefont{Sopova}} \bibnamefont{and}
  \bibinfo{author}{\bibfnamefont{L.~H.} \bibnamefont{Ford}}
  (\bibinfo{year}{2005}), \eprint[http://arXiv.org/abs]{quant-ph/0504143}.

\bibitem[{\citenamefont{Sopova and Ford}(2002)}]{Sopova:2002cs}
\bibinfo{author}{\bibfnamefont{V.}~\bibnamefont{Sopova}} \bibnamefont{and}
  \bibinfo{author}{\bibfnamefont{L.~H.} \bibnamefont{Ford}},
  \bibinfo{journal}{Phys. Rev.} \textbf{\bibinfo{volume}{D66}},
  \bibinfo{pages}{045026} (\bibinfo{year}{2002}),
  \eprint[http://arXiv.org/abs]{quant-ph/0204125}.

\bibitem[{\citenamefont{Klinkhammer}(1991)}]{Klinkhammer}
\bibinfo{author}{\bibfnamefont{G.}~\bibnamefont{Klinkhammer}},
  \bibinfo{journal}{Phys. Rev. D} \textbf{\bibinfo{volume}{43}},
  \bibinfo{pages}{2542} (\bibinfo{year}{1991}).

\bibitem[{\citenamefont{Folacci}(1992)}]{Folacci:1992xg}
\bibinfo{author}{\bibfnamefont{A.}~\bibnamefont{Folacci}},
  \bibinfo{journal}{Phys. Rev.} \textbf{\bibinfo{volume}{D46}},
  \bibinfo{pages}{2726} (\bibinfo{year}{1992}).

\bibitem[{\citenamefont{Borde et~al.}(2002)\citenamefont{Borde, Ford, and
  Roman}}]{Borde01}
\bibinfo{author}{\bibfnamefont{A.}~\bibnamefont{Borde}},
  \bibinfo{author}{\bibfnamefont{L.}~\bibnamefont{Ford}}, \bibnamefont{and}
  \bibinfo{author}{\bibfnamefont{T.~A.} \bibnamefont{Roman}},
  \bibinfo{journal}{Phys. Rev. D} \textbf{\bibinfo{volume}{65}},
  \bibinfo{pages}{084002} (\bibinfo{year}{2002}),
  \eprint[http://arXiv.org/abs]{gr-qc/109061}.

\bibitem[{\citenamefont{Ford and Roman}(1995)}]{F&Ro95}
\bibinfo{author}{\bibfnamefont{L.~H.} \bibnamefont{Ford}} \bibnamefont{and}
  \bibinfo{author}{\bibfnamefont{T.~A.} \bibnamefont{Roman}},
  \bibinfo{journal}{Phys. Rev. D} \textbf{\bibinfo{volume}{51}},
  \bibinfo{pages}{4277} (\bibinfo{year}{1995}), \bibinfo{note}{gr-qc/9410043}.

\bibitem[{\citenamefont{Ford and Roman}(1996)}]{F&Ro96b}
\bibinfo{author}{\bibfnamefont{L.~H.} \bibnamefont{Ford}} \bibnamefont{and}
  \bibinfo{author}{\bibfnamefont{T.~A.} \bibnamefont{Roman}},
  \bibinfo{journal}{Phys. Rev. D} \textbf{\bibinfo{volume}{53}},
  \bibinfo{pages}{1988} (\bibinfo{year}{1996}).

\bibitem[{\citenamefont{Bordag and Lindig}(1996)}]{Bordag:1996}
\bibinfo{author}{\bibfnamefont{M.}~\bibnamefont{Bordag}} \bibnamefont{and}
  \bibinfo{author}{\bibfnamefont{J.}~\bibnamefont{Lindig}},
  \bibinfo{journal}{J. Phys.} \textbf{\bibinfo{volume}{A29}},
  \bibinfo{pages}{4481} (\bibinfo{year}{1996}).

\bibitem[{\citenamefont{Saharian}(2001)}]{Saharian:2000mw}
\bibinfo{author}{\bibfnamefont{A.~A.} \bibnamefont{Saharian}},
  \bibinfo{journal}{Phys. Rev.} \textbf{\bibinfo{volume}{D63}},
  \bibinfo{pages}{125007} (\bibinfo{year}{2001}), \eprint{hep-th/0012185}.

\bibitem[{\citenamefont{Graham et~al.}(2002)\citenamefont{Graham, Jaffe,
  Khemani, Quandt, Scandurra, and Weigel}}]{Graham:2002cas}
\bibinfo{author}{\bibfnamefont{N.}~\bibnamefont{Graham}},
  \bibinfo{author}{\bibfnamefont{R.~L.} \bibnamefont{Jaffe}},
  \bibinfo{author}{\bibfnamefont{V.}~\bibnamefont{Khemani}},
  \bibinfo{author}{\bibfnamefont{M.}~\bibnamefont{Quandt}},
  \bibinfo{author}{\bibfnamefont{M.}~\bibnamefont{Scandurra}},
  \bibnamefont{and} \bibinfo{author}{\bibfnamefont{H.}~\bibnamefont{Weigel}},
  \bibinfo{journal}{Nucl. Phys.} \textbf{\bibinfo{volume}{B645}},
  \bibinfo{pages}{49} (\bibinfo{year}{2002}),
  \eprint[http://arXiv.org/abs]{hep-th/0207120}.

\bibitem[{\citenamefont{Abramowitz and Stegun}(1972)}]{Abramowitz}
\bibinfo{author}{\bibfnamefont{M.}~\bibnamefont{Abramowitz}} \bibnamefont{and}
  \bibinfo{author}{\bibfnamefont{I.~A.} \bibnamefont{Stegun}},
  \emph{\bibinfo{title}{Handbook of Mathematical Functions With Formulas,
  Graphs, and Mathematical Tables}} (\bibinfo{publisher}{U.S. government
  printing office}, \bibinfo{address}{Washington}, \bibinfo{year}{1972}).

\bibitem[{\citenamefont{Morse and Feshbach}(1953)}]{MorseFeshbach}
\bibinfo{author}{\bibfnamefont{P.~M.} \bibnamefont{Morse}} \bibnamefont{and}
  \bibinfo{author}{\bibfnamefont{H.}~\bibnamefont{Feshbach}},
  \emph{\bibinfo{title}{Methods of theoretical physics}}
  (\bibinfo{publisher}{McGraw-Hill}, \bibinfo{address}{New York},
  \bibinfo{year}{1953}).

\bibitem[{\citenamefont{Meixner and {Sch\"afke}}(1954)}]{Meixner}
\bibinfo{author}{\bibfnamefont{J.~W.} \bibnamefont{Meixner}} \bibnamefont{and}
  \bibinfo{author}{\bibfnamefont{R.~W.} \bibnamefont{{Sch\"afke}}},
  \emph{\bibinfo{title}{Mathieusche Funktionen und {Sph\"aroidfunktionen}}}
  (\bibinfo{publisher}{Springer-Verlag}, \bibinfo{address}{Berlin},
  \bibinfo{year}{1954}).

\bibitem[{\citenamefont{Alhargan}(2000{\natexlab{a}})}]{Alhargan:2000}
\bibinfo{author}{\bibfnamefont{F.}~\bibnamefont{Alhargan}},
  \bibinfo{journal}{ACM Transactions on Mathematical Software}
  \textbf{\bibinfo{volume}{26}}, \bibinfo{pages}{390}
  (\bibinfo{year}{2000}{\natexlab{a}}).

\bibitem[{\citenamefont{Alhargan}(2000{\natexlab{b}})}]{Alhargan:2000a}
\bibinfo{author}{\bibfnamefont{F.}~\bibnamefont{Alhargan}},
  \bibinfo{journal}{ACM Transactions on Mathematical Software}
  \textbf{\bibinfo{volume}{26}}, \bibinfo{pages}{408}
  (\bibinfo{year}{2000}{\natexlab{b}}).

\bibitem[{\citenamefont{Falloon et~al.}(2003)\citenamefont{Falloon, Abbott, and
  Wang}}]{Falloon:2002}
\bibinfo{author}{\bibfnamefont{P.~E.} \bibnamefont{Falloon}},
  \bibinfo{author}{\bibfnamefont{P.~C.} \bibnamefont{Abbott}},
  \bibnamefont{and} \bibinfo{author}{\bibfnamefont{J.~B.} \bibnamefont{Wang}},
  \bibinfo{journal}{Journal of Physics A: Math. Gen.}
  \textbf{\bibinfo{volume}{36}}, \bibinfo{pages}{5477} (\bibinfo{year}{2003}),
  \eprint[http://arXiv.org/abs]{math-ph/0212051}.

\bibitem[{\citenamefont{Ford and Svaiter}(1998)}]{Ford:1998he}
\bibinfo{author}{\bibfnamefont{L.~H.} \bibnamefont{Ford}} \bibnamefont{and}
  \bibinfo{author}{\bibfnamefont{N.~F.} \bibnamefont{Svaiter}},
  \bibinfo{journal}{Phys. Rev.} \textbf{\bibinfo{volume}{D58}},
  \bibinfo{pages}{065007} (\bibinfo{year}{1998}), \eprint{quant-ph/9804056}.

\end{thebibliography}

\end{document}